\documentclass[aps,prb,reprint,longbibliography,superscriptaddress]{revtex4-2}
\usepackage{mathrsfs}
\usepackage{amsmath,gensymb}
\usepackage{amsfonts}
\usepackage{amssymb}
\usepackage{amsthm}
\usepackage{graphicx}
\usepackage{natbib}
\usepackage{color}
\usepackage{hyperref}
\usepackage{bm}
\usepackage[caption=false]{subfig}
\usepackage{verbatim}
\usepackage[normalem]{ulem}
\usepackage{soul}

\begin{document}
\title{Gate-tunable nonlocal Josephson effect through magnetic van der Waals bilayers}

\author{G. A. Bobkov}
\affiliation{Moscow Institute of Physics and Technology, Dolgoprudny, 141700 Moscow Region, Russia}

\author{D.S. Rabinovich}
\affiliation{Skolkovo Institute of Science and Technology, 121205 Moscow, Russia}
\affiliation{Moscow Institute of Physics and Technology, Dolgoprudny, 141700 Moscow region, Russia}

\author{A. M. Bobkov}
\affiliation{Moscow Institute of Physics and Technology, Dolgoprudny, 141700 Moscow region, Russia}

\author{I.V. Bobkova}
\affiliation{Moscow Institute of Physics and Technology, Dolgoprudny, 141700 Moscow region, Russia}
\affiliation{National Research University Higher School of Economics, 101000 Moscow, Russia}

\begin{abstract}
It is well-known that the proximity effect at superconductor/ferromagnet (S/F) interfaces produces damped oscillatory behavior of the Cooper pair wave function within the ferromagnetic regions, which is analogous to the inhomogeneous Fulde-Ferrell-Larkin-Ovchinnikov (FFLO) superconductivity. It is often called the mesoscopic FFLO state and gives rise to $0-\pi$-transitions in S/F/S Josephson junctions. This paper offers an analysis of the proximity effect at interfaces between superconductors and magnetic van der Waals (vdW) bilayers.  The specific feature of the proximity effect in the vdW bilayer systems is the presence of non-local Cooper pairs. We predict that the mesoscopic FFLO state formed by such pairs is sensitive to the difference
between on-site energies of the monolayers composing the
bilayer and, thus, can be controlled by applying a gating
potential to one of the monolayers. This opens the possibility of implementing gate-controlled $0-\pi$ transitions
in Josephson junctions through the magnetic vdW bilayer weak links.
\end{abstract}

\maketitle

\section{Introduction}

The proximity effect between different materials modifies their properties near the interface. In the case of a superconductor/normal metal (S/N) interface, the Cooper pairs can penetrate into the normal metal at a typical distance of the order of the normal coherence length $\xi_N$. In the diffusive case this length is of the order of $\xi_N \sim \sqrt{D/2\pi T}$, where $D$ is the diffusion constant and $T$ is the temperature. In the ballistic case $\xi_N \sim v_F/2\pi T$, where $v_F$ is the electron Fermi velocity in the normal metal. At superconductor/ferromagnet (S/F) interfaces the singlet Cooper pairs penetrating into the ferromagnetic region are partially converted into triplet ones and acquire finite momentum. This is caused by the presence of the exchange field $h$ in the ferromagnet. Because of this field, the energy of a spin-up electron decreases by $h$, while the energy of a spin-down electron increases by $h$. To compensate this energy difference, the kinetic energy of the spin-up (down) electron increases (decreases), what gives rise to the difference of the absolute values of their momenta \cite{Demler1997,Buzdin2005,Bergeret2005}. Due to the finite momentum the decay of the singlet and triplet pairs into the ferromagnet is superimposed by oscillations. The finite momentum of the pair and the resulting spatial oscillations of its wave function are the key features of the finite-momentum inhomogeneous superconductivity, predicted earlier by Fulde and Ferrell \cite{Fulde1964} and  Larkin and Ovchinnikov\cite{Larkin1964}. Thus, the proximity-induced superconducting correlations in ferromagnets are often called the mesoscopic Fulde-Ferrel-Larkin-Ovchinnikov (FFLO) state \cite{Buzdin2005} in the literature.

In the ballistic case the scale of the spatial oscillations of the proximity-induced pair  correlations in the ferromagnet is $\sim v_F/2h$ if $h \gg T$ \cite{Buzdin1982,Radovic2003}. In the diffusive case the penetration length and the period of oscillations are of the same order of the magnetic coherence length $\xi_F \sim \sqrt{D/h} \ll \xi_N$. The presence of inhomogeneous magnetization in the interlayer region essentially modifies the triplet proximity effect in ferromagnetic interlayers of JJs giving rise to the long-range range triplet component, which does not oscillate and decays at $\xi_N$ \cite{Bergeret2005,Eschrig2015,Linder2015}.

The oscillations of the proximity-induced singlet superconducting correlations in the ferromagnet lead to $0-\pi$ transitions of  the ground state superconducting phase difference in superconductor/ferromagnet/superconductor Josephson junctions (S/F/S JJs) upon varying the length of the ferromagnetic weak link \cite{Golubov2004,Buzdin2005}. Spatial oscillations of proximity-induced triplet superconductivity were also predicted at superconductor/antiferromagnet (S/AF) interfaces \cite{Bobkov2023}. Yet the physical mechanism is different from S/F heterostructures and does not lead to oscillations of the singlet component. Thus, it does not result in $0-\pi$ transitions in S/AF/S JJs. At the same time, $0-\pi$ transitions in S/AF/S JJs upon atomic variations of the AF interlayer length were predicted \cite{Andersen2006,Enoksen2013,Bulaevskii2017}. The physical reason for such $0-\pi$ transitions is related to the reconstruction of the Andreev bound states spectra inside the interlayer and thus atomically sharp S/AF interfaces are required to observe the effect. 

JJs with the $\pi$ ground state phase difference ($\pi$-JJs) are implemented as
$\pi$-phase shifters in superconducting and quantum electronics \cite{Yamashita2005,Feofanov2010,Shcherbakova2015}. The broad application prospects motivated an active search for novel principles and possibilities  of external control of the switching between the $0$ and $\pi$ ground states. Various suggestions of externally controlled $0-\pi$ transitions in Josephson junctions include the temperature induced $0-\pi$ transitions \cite{Ryazanov2001},   $0-\pi$ transitions induced by the electrostatic gating \cite{vanDam2006, jorgensen2007critical} in quantum dot JJs, as well as by spin-independent \cite{Volkov1995,Morpurgo1998,Baselmans1999,Huang2002,Yip2000,Heikkila2000,Wilheim1998,golikova2021controllable} and spin-dependent \cite{Bobkova2010,Bobkov2011} nonequilibrium quasiparticle distribution, via the manipulation of the mutual orientation of the ferromagnets of a spin-valve inside a JJ \cite{Gingrich2016} and  by controlling the exchange field orientation in S/F/S JJs with spin-orbit coupling \cite{Bujnowski2019,Eskilt2019}.
In a two-dimensional electron gas (2DEG) $0-\pi$ transitions can also be controlled by an applied magnetic field or gating \cite{ke2019ballistic}. The supercurrent-induced long-range triplet component in Josephson junctions \cite{Silaev2020}
provides a possibility of a supercurrent-controllable $0-\pi$
transition \cite{Mazanik2021}.

The recent discoveries of van der Waals (vdW) superconducting and magnetic materials \cite{Staley2009,Deng2018,Gong2017,Huang2017,Gibertini2019,Liu2016,Novoselov2016} have provided a new platform to realize JJs demonstrating $0-\pi$ transitions with atomically flat and magnetically
sharp interfaces and atomically uniform thickness \cite{Ai2021,Dvir2021,Kang2022,Qiu2023}.  This type of functional nanodevices has great potential for superconducting spintronics, 
quantum circuits and cryogenic memories.

In this paper we consider the penetration of proximity-induced superconducting correlations into bilayer magnetic vdW materials or heterostructures. It is demonstrated that at low electron transparency (interlayer hopping) between the interacting vdW layers composing the bilayer the proximity-induced Cooper pairs can be divided into local and nonlocal ones. The penetration of the nonlocal pairs into nonsuperconducting materials substantially differs from the penetration of the local pairs. In particular, the length of spatial oscillations of such pairs can be tuned by applying the gating potential to one of the vdW layers.  It provides a way to identify the contribution of such pairs into the Josephson current and to generate $0-\pi$ transitions via gating. It is worth noting that the possibility to control $0-\pi$ transitions via gating results from the unique sensitivity of spatial behavior of the nonlocal pairs to this parameter and is not related to any tuning of the electron density as in experiments with quantum dots and 2DEG \cite{vanDam2006, jorgensen2007critical,ke2019ballistic}.  Further we demonstrate that at higher values of the interlayer hopping elements there is a region of strong hybridization between the local and nonlocal pairs, where the physical properties of all pairs differ from the conventional bulk behavior and the corresponding oscillation periods can be controlled by gating, which again gives rise to unusual and tunable behavior of the Josephson current. We investigate and compare manifestations of the nonlocal proximity-induced superconducting pairs in magnetic bilayers with antiferromagnetic, ferrimagnetic and ferromagnetic interlayer ordering. We also note that although proximity effects in vdW and layered S/F, S/F/S, F/S/F and S/F/F heterostructures with atomic thicknesses are being studied very actively \cite{Aikebaier2022,Kang2021,Wickramaratne2021,Jo2023,Jiang2020,Ai2021,Idzuchi2021,Buzdin2003,Tollis2005,Montiel2009,Devizorova2017,Devizorova2019,Bobkov2024_vdW,Bobkov2024_spin,Ianovskaia2024}, the majority of those works consider a sandwich geometry, where the system is translational invariant along the interface and contains a small number of atomic layers, which does not allow one to observe a spatial dependence of the superconducting properties. In contrast to the previous studies, we focus on the spatial behavior of the proximity-induced superconducting correlations in the plane of the layers and its dependence on the applied gating potential.

The paper is organized as follows. In Sec.~\ref{system} the system under consideration is described. Sec.~\ref{nonlocal} is devoted to analytical investigation of different proximity-induced Cooper pairs and their penetration into non-superconducting bilayers: in Sec.~\ref{Eilenberger} the generalized quasiclassical Eilenberger equation for the description of bilayer vdW heterostructures is derived and in Sec.~\ref{pairs} this equation is applied to the analytical description  of proximity-induced Cooper pairs. In Sec.~\ref{Josephson}  results for the Josephson current through magnetic vdW bilayers are presented: in Sec.~\ref{nonlocal_JJ} the contribution of the nonlocal pairs in JJs with AF bilayer weak links is discussed; Sec.~\ref{zero_pi} demonstrates how the nonlocal pair contribution gives rises to $0-\pi$ transitions induced by gating and in Sec.~\ref{ferromagnet_JJ} we consider how the nonlocal pairs manifest themselves in JJs with ferromagnet/normal metal interlayers. Our conclusions are formulated in Sec.~\ref{conclusions}.

\section{System and model}

\label{system}

\begin{figure}[tb]
	\begin{center}
		\includegraphics[width=85mm]{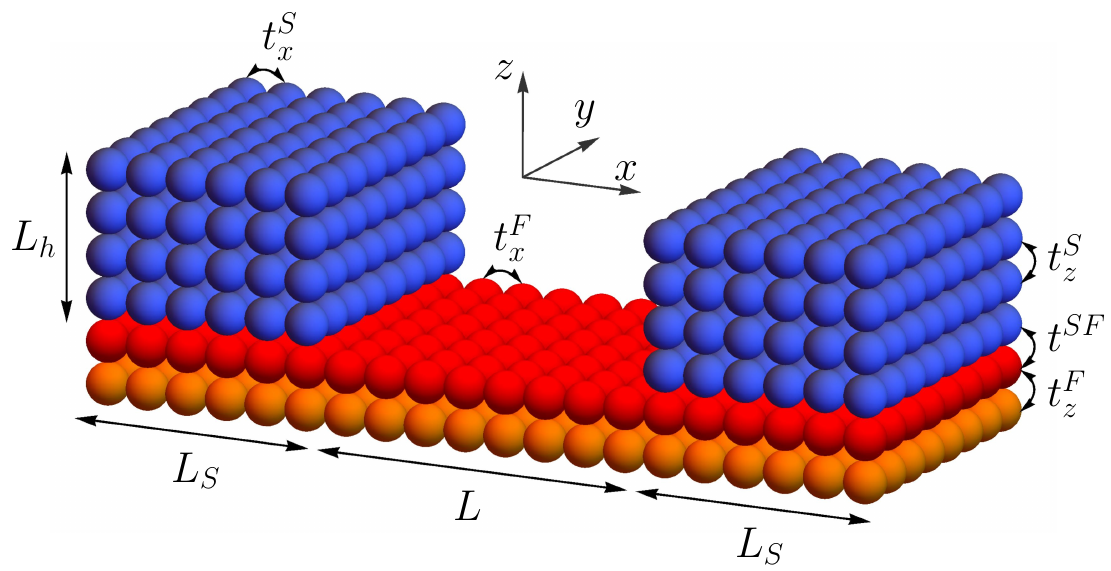}
\caption{Sketch of the considered JJ. Superconducting sites are depicted by blue, while red and orange layers represent the upper ferromagnetic layer (F1) and bottom ferromagnetic layer (F2), respectively. The interlayer magnetic ordering can be of different types: AF, F or ferrimagnetic, see text for more details. Different hopping elements are also shown.}
 \label{fig:sketch}
	\end{center}
 \end{figure}

The system under consideration is shown in Fig.~\ref{fig:sketch}. It represents a planar JJ made of layered superconducting leads and a bilayer magnetic weak link. The system is modelled by the following tight-binding Hamiltonian on a square lattice:
\begin{align}
\hat{H} &=  -  \sum\limits_{<\bm i \bm j>, \sigma} t_{\bm i \bm j} c^{\dagger}_{\bm i \sigma} c_{\bm j \sigma} + \sum\limits_{\bm i, \alpha \beta}  c^{\dagger}_{\bm i \alpha} (\bm h_{\bm i} \bm \sigma)_{\alpha \beta} c_{\bm i \beta}- \nonumber \\
&\sum\limits_{\bm i \sigma} \mu_{\bm i} c^{\dagger}_{\bm i \sigma} c_{\bm i \sigma} + 
\sum\limits_{\bm i} (\Delta_{\bm i}  c^{\dagger}_{\bm i \uparrow} c^{\dagger}_{\bm i \downarrow} + h.c.) 
\label{eq:hamiltonian}
\end{align}
Here $c_{\bm i,\sigma}$ is an annihilation operator for electrons at site $\bm i$ and for spin $\sigma = \uparrow, \downarrow$. $\mu_{\bm i}$ is the on-site energy, which is assumed to be $\mu_{\bm i} = \mu_{1,2}$ at sites belonging to the upper (F1) and lower (F2) ferromagnetic layers, respectively, and $\mu_{\bm i} = \mu_S$ at the sites belonging to the superconducting leads. The on-site energy  determines the filling factors of the conduction band of the materials and in the case of isolated layers correspond to their chemical potentials. We assume only nearest-neighbor hopping with $t_{\bm i \bm j} = t_{x}^F$ being the intralayer hopping element for both ferromagnetic layers and $t_{\bm i \bm j} = t_{z}^F$  being the hopping element between the F1 and F2 layers. Anisotropic intralayer (interlayer) hopping elements between nearest superconducting sites are denoted by $t_{\bm i \bm j} = t_{x(z)}^S$ and $t_{\bm i \bm j} = t^{SF}$ is the hopping element between the upper ferromagnetic layer and the superconducting leads. $\langle \bm i \bm j \rangle$ means summation over nearest neighbors. $h.c.$ means hermitian conjugate. 

$\bm h_{\bm i}$ is the exchange field in the  ferromagnetic layers, which is equal to $\bm h_{1,2}$ in the F1 and F2 layers, respectively and is assumed to be spatially homogeneous in the plane of each layer. In superconducting leads $\bm h_{\bm i} = 0$. Below we consider different physical cases: $\bm h_1 = \bm h_2$ (ferromagnetic interlayer ordering), $\bm h_1 = -\bm h_2$ (antiferromagnetic interlayer ordering), $\bm h_1$ is opposite to $\bm h_2$ and  $|\bm h_1| \neq |\bm h_2|$ (ferrimagnetic interlayer ordering). The case $\bm h_1 \neq 0$ and $\bm h_2 = 0$ modeling a bilayer consisting of a ferromagnetic and a paramagnetic layers is also considered. Such a bilayer structure can be implemented by exfoliation of vdW magnetic materials \cite{Tan2021,Lu2023,Seo2021} or using chemical vapor deposition (CVD) processes. Thin films of some intermetallic materials, such as $\mathrm{GdIr_2Si_2}$ or $\mathrm{TbRh_2Si_2}$, seem to be good candidates \cite{Usachev2020,Schultz2021}. Also the bilayers can be assembled from magnetic and nonmagnetic monolayers of different vdW materials \cite{Liu2016}, possibly with an insulating vdW layer between them in order to exclude the direct exchange interaction and to control the magnetization orientations in the layers by the applied magnetic field. 

$\Delta_{\bm i}$ is the superconducting order parameter in the superconducting leads, which is zero at sites belonging to the magnetic layers. In principle, it should be calculated self-consistently as $\Delta_{\bm i} = \lambda \langle c_{\bm i \downarrow} c_{\bm i \uparrow} \rangle$, where $\lambda$ is the pairing constant. If the height of the leads $L_h$ is less than the superconducting coherence length $\xi_S$, the superconducting order parameter is approximately spatially constant in the leads, although its value can be suppressed with respect to the bulk value due to the proximity to the ferromagnetic layer. When calculating and analyzing the Josephson current we are not interested in the exact value of this constant. For this reason here the order parameter in the leads is not calculated self-consistently. The system is assumed to be in the ballistic limit, that is the impurity scattering is absent.

\section{Penetration of nonlocal correlations into nonsuperconducting regions}

\label{nonlocal}

\begin{figure*}[tb]
	\begin{center}
		\includegraphics[width=170mm]{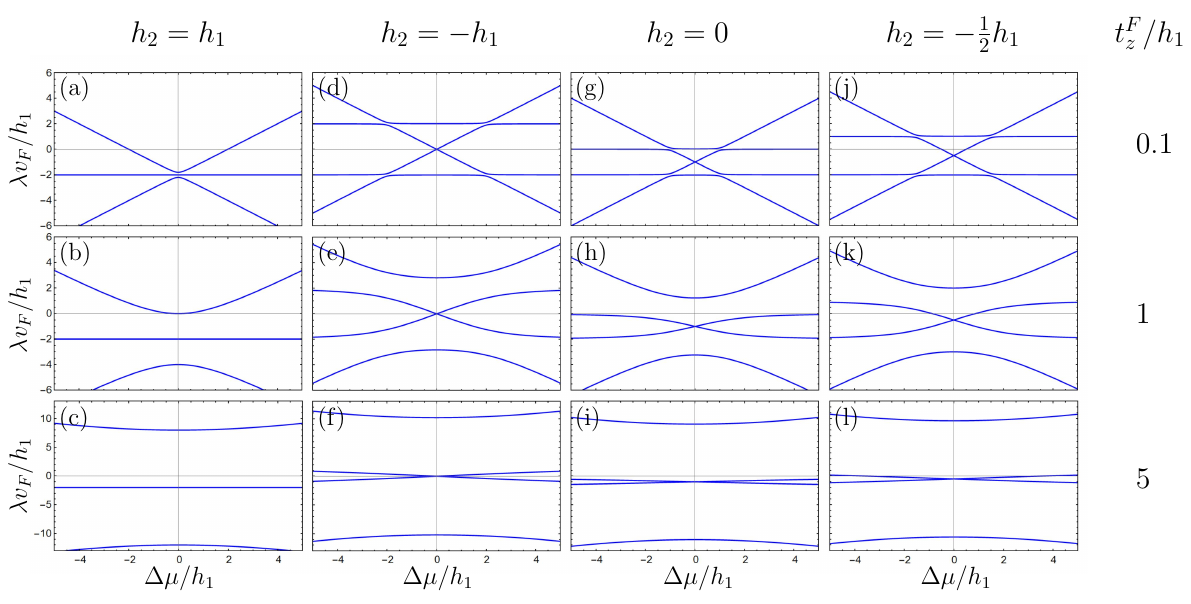}
\caption{Eigenvalues $\lambda_k$, which determine the inverse oscillation periods of the four possible solutions of the anomalous Green's function in a magnetic bilayer as functions of $\Delta \mu = \mu_1 - \mu_2$. Different columns correspond to different types of magnetic bilayers: ferromagnetic [ panels (a)-(c)], antiferromagnetic [(d)-(f)], bilayer consisting of a ferromagnetic and a paramagnetic layer [(g)-(i)] and a ferrimagnetic bilayer [(j)-(l)]. Different horizontal rows correspond to different values of the  interlayer hopping $t_z^F$.}
 \label{fig:oscillations}
	\end{center}
 \end{figure*}

Here we consider a non-superconducting bilayer, which has an interface with a superconductor and investigate the proximity penetration of the Cooper pairs from the superconductor  into the depth of the bilayer.  Details of the geometry of the interface do not matter in this consideration. The superconductor can be located on top of a part of the bilayer (a half of the sketch presented in Fig.~\ref{fig:sketch} without the second lead) or the superconductor can have a contact with both non-superconducting layers. These details only influence the amplitudes of the different types of Cooper pairs induced in the bilayer by the proximity effect, yet do not affect their spatial behavior inside the bilayer, penetration depth and oscillating period. The problem can be studied analytically in the framework of the Eilenberger equation for the quasiclassical Green's function describing electronic behavior in the bilayer.

\subsection{Derivation of the two-layer quasiclassical eilenberger equation}

\label{Eilenberger}

At first we describe the main steps of the derivation of the quasiclassical Eilenberger equation. We begin with the derivation of the Gor'kov equation for the full quantum mechanical Green's function. Let us consider a general bilayer system [not the particular planar JJ described by the Hamiltonian (\ref{eq:hamiltonian})] described by the following tight-binding Hamiltonian: 
\begin{align}
\hat H=\sum\limits_{\bm i,\alpha,\beta}\hat c^\dagger_{\bm i,\alpha} (\bm {\hat h}\bm \sigma)_{\alpha\beta} \hat c_{\bm i,\beta}-\sum\limits_{\bm i,\sigma}\hat c^\dagger_{\bm i,\sigma}\hat \mu \hat c_{\bm i,\sigma}-\sum\limits_{<\bm i\bm j>,\sigma}\hat c^\dagger_{\bm i,\sigma} t \hat c_{\bm j,\sigma} \nonumber \\
-\sum\limits_{\bm i,\sigma}\hat c^\dagger_{\bm i,\sigma}\rho_x t_{12}\hat c_{\bm i,\sigma} +\sum\limits_{\bm i}\left[\hat c_{\bm i,\uparrow} \hat \Delta_{\bm i}\hat c_{\bm i,\downarrow}+h.c.\right]~~~~~~~~~
\label{eq:hamiltonian_gen}
\end{align}
Here $\hat c_{\bm i,\sigma} = (c_{\bm i,\sigma}^{1}, c_{\bm i,\sigma}^{2})^T$ is a vector composed of annihilation operators for electrons belonging to the first and the second layers at site $\bm i$ in the plane of each layer and for spin $\sigma = \uparrow, \downarrow$. $\hat \mu=\left(\begin{matrix}\mu_1&0\\0&\mu_2\\\end{matrix}\right)$ is the matrix composed of on-site energies of the first and the second layers, respectively. $t$ is the hopping element in plane of each layer and $t_{12}$ is the hopping element between the first and the second layers. $\bm {\hat h}=\left(\begin{matrix}\bm h_1&0\\0&\bm h_2\\\end{matrix}\right)$ is the matrix containing the exchange field of both layers and $\hat \Delta_{\bm i}=\left(\begin{matrix} \Delta_{1,\bm i}&0\\0& \Delta_{2, \bm i}\\\end{matrix}\right)$ is the matrix composed of superconducting order parameters of the layers, which vanishes in the nonsuperconducting layer. 

The Matsubara Green's function in the two-layer formalism is an $8 \times 8$ matrix in the direct product of spin, particle-hole and layer spaces. Introducing the two-layer Nambu spinor $\check \psi_{\bm i} = (\hat c_{{\bm i},\uparrow}^1, \hat c_{\bm i,\downarrow}^1, \hat c_{\bm i,\uparrow}^2,\hat c_{\bm i,\downarrow}^2, \hat c_{\bm i,\uparrow}^{1\dagger}, \hat c_{\bm i,\downarrow}^{1\dagger}, \hat c_{\bm i,\uparrow}^{2\dagger}, \hat c_{\bm i,\downarrow}^{2\dagger})^T$ we define the Green's function as follows: 
\begin{eqnarray}
\check G_{\bm i \bm j}(\tau_1, \tau_2) = -\langle T_\tau \check \psi_{\bm i}(\tau_1) \check \psi_{\bm j}^\dagger(\tau_2) \rangle,
\label{Green_Gorkov}
\end{eqnarray}
where $\langle T_\tau ... \rangle$ means  imaginary time-ordered thermal averaging. Introducing Pauli matrices in spin, particle-hole and layer spaces, as $\sigma_k$, $\tau_k$ and $\rho_k$ ($k=0,x,y,z$) and operator $\hat j $ as
\begin{eqnarray}
\hat j \hat c_{\bm i} = \sum\limits_{<\bm i\bm j>}t\hat c_{\bm j}
\label{op_j}
\end{eqnarray}
and expanding $\check G_{\bm i \bm j}(\tau_1 - \tau_2) = T \sum \limits_{\omega_m} e^{-i \omega_m (\tau_1 - \tau_2)} \check G_{\bm i \bm j}(\omega_m)$ over Matsubara frequencies $\omega_m = \pi T(2m+1)$, one can obtain the Gor'kov equation for Green's function $\check G_{\bm i \bm j}(\omega_m)$. The derivation is similar to that described in Ref.~\onlinecite{Ianovskaia2024}. The resulting Gor'kov equation takes the form:
\begin{align}
G_{\bm i}^{-1} \check G_{\bm i \bm j}(\omega_m) = \delta_{\bm i \bm j}, \label{gorkov_eq_ml}
\end{align}
\begin{align}
G_{\bm i}^{-1} = \tau_z \left( \hat j + \hat \mu - \check \Delta_{\bm i} i \sigma_y - \bm {\hat h} \check {\bm \sigma}   + t_{12}\rho_x\right) + i \omega_m .
\label{G_i}
\end{align}
where $\check \Delta_{\bm i} = \hat \Delta_{\bm i} \tau_+ + \hat\Delta_{\bm i}^* \tau_-$ with $\tau_\pm  = (\tau_x \pm i \tau_y)/2$ and $\check {\bm \sigma} = \bm \sigma (1+\tau_z)/2 + \bm \sigma^* (1-\tau_z)/2$ is the quasiparticle spin operator. Further we consider the Green's function in the mixed representation:
\begin{eqnarray}
\check G(\bm R, \bm p) = F(\check G_{\bm i \bm j}) = \int d^2 r e^{-i \bm p(\bm i - \bm j)}\check G_{\bm i \bm j},
\label{mixed}
\end{eqnarray}
where $\bm R=(\bm i+\bm j)/2$ and the integration is over $\bm r = \bm i - \bm j$. To further simplify calculations and to present the Gor'kov equation in a more common form we define the following  transformed Green's function:

\begin{eqnarray}
\check {\tilde G}(\bm R, \bm p) = 
\left(
\begin{array}{cc}
1 & 0 \\
0 & -i\sigma_y
\end{array}
\right)_\tau  \check G(\bm R, \bm p)  
\left(
\begin{array}{cc}
1 & 0 \\
0 & -i\sigma_y
\end{array}
\right)_\tau ,~~~~
\label{unitary}
\end{eqnarray}
where and below subscript $\tau$ means that the explicit matrix structure corresponds to the particle-hole  space. In the quasiclassical approximation we disregard terms of the order of $1/(\xi_S p_F) \ll 1$ with respect to unity, where $p_F$ is the Fermi momentum, $\xi_S = v_F/|\Delta|$ is the superconducting coherence length and $\hbar=1$. For this reason we can approximate $\hat \Delta_{\bm i} \approx \hat \Delta(\bm R)$. The neglected terms $\sim (\bm \nabla \hat \Delta_{\bm i}) \bm r \sim |\Delta_{1,2}|/(p_F \xi_S)$. In general, the same is valid for the exchange field if it is inhomogeneous in the plane of the layer. With these approximations we obtain:
\begin{align}
    &\left( 
    i\omega_m\tau_z+F(\hat j \check {\tilde G})  +\hat \mu+\tau_z\check \Delta (\bm R)-\bm {\hat h}\bm \sigma\tau_z+t_{12}\rho_x
    \right) \times \nonumber \\
    &\check {\tilde G}(\bm R, \bm p)=1
    \label{eilenberger_left}
\end{align}
Analogously, by deriving the Gor'kov equation, where the operator $G^{-1}$ acts on the Green’s function from the right, one can obtain:
\begin{align}
    \check {\tilde G}(\bm R, \bm p)\left( i\omega_m\tau_z+F( \check {\tilde G}\hat j)  +\hat \mu+ \right. \nonumber \\
   \left. \tau_z\check \Delta (\bm R)-\bm {\hat h}\bm \sigma\tau_z+t_{12}\rho_x
    \right) =1
    \label{eilenberger_right}
\end{align}
Subtracting Eqs.~(\ref{eilenberger_left}) and (\ref{eilenberger_right}) we come to the following equation:
\begin{align}
&\left[i \omega_m \tau_z + \hat \mu + \tau_z \check \Delta - \bm {\hat h} \bm \sigma \tau_z \rho_z +t_{12}\rho_x , \check {\tilde G}(\bm R, \bm p) \right] +  \nonumber \\
&F([\hat j, \check {\hat G}])= 0
\label{eilenberger_sub}
\end{align}
We introduce a quasiclassical $\xi$-integrated Green's function:
\begin{eqnarray}
\check g(\bm R, \bm p_F) = -\frac{1}{i \pi} \int \check {\tilde G}(\bm R, \bm p)d\xi,
\label{quasi_green} 
\end{eqnarray}
where  $a$ is the lattice constant and the integration is taken over the electron energy $ \xi=-2t [\cos(p_x a)+\cos(p_y a)]$ around the Fermi surface.  The resulting quasiclassical Green's function $\hat g$ only depends on the center of mass coordinate $\bm R = (\bm i + \bm j)/2$ of the paired electrons with momenta $\bm p_F$ and $-\bm p_F$ at the Fermi surface and the direction of the trajectory determined by $\bm p_F$. After the $\xi$-integration of Eq.~(\ref{eilenberger_sub}) the Eilenberger equation for the quasiclassical Green's function takes the form:
\begin{align}\nonumber
    \left[i \omega_m \tau_z + \hat \mu + \tau_z \check \Delta - \bm {\hat h} \bm \sigma \tau_z \rho_z +t_{12}\rho_x , \check g(\bm R, \bm p_F) \right] + \\ 
    +i\bm v_F \bm \nabla \check g(\bm R,\bm p_F) = 0, 
\label{eilenberger_ballistic}
\end{align}
where $\bm v_F = d\xi/d \bm p|_{\bm p = \bm p_F} = 2ta(\sin [p_{F,x}a],\sin [p_{F,y}a]).$ Eq.~(\ref{eilenberger_ballistic}) should be supplemented by the normalization condition
\begin{eqnarray}
\check g^2(\bm R, \bm p_F) = 1 .
\label{norm}
\end{eqnarray}

\subsection{Proximity-induced superconducting correlations in a non-superconducting bilayer}

\label{pairs}

Let us apply the derived Eilenberger equation to the non-superconducting bilayer, which serves as an interlayer of the considered JJ. Then  the order parameter $\Delta_{\bm i}=0$ in both layers, $t_{12} \equiv t_z^F$ according to the Hamiltonian (\ref{eq:hamiltonian}) and the generalized Eilenberger equation takes the form: 
\begin{align}\nonumber
    \left[i \omega_m \tau_z + \hat \mu  - \bm {\hat h} \bm \sigma \tau_z \rho_z +t_{z}^F\rho_x , \check g(\bm R, \bm p_F) \right] + \\ 
    +i\bm v_F \bm \nabla \check g(\bm R,\bm p_F) = 0, 
\label{eilenberger_ballistic_main}
\end{align}
We work at temperatures close to the critical temperature and linearize the Eilenberger equation with respect to the anomalous component of the Green's function, which is assumed to be small with respect to unity. This approximation is valid if the temperature is close to the critical temperature of the superconducting leads or if the interfaces between the superconductors and the bilayer are low-transparent. In this approximation the Green's function takes the form:
\begin{equation}
\check g = 
\left(
\begin{array}{cc}
\hat g_N & \hat f \\
\hat {\tilde f} & \hat {\tilde g}_N
\end{array}
\right)_{\tau},
\label{gf_linearized}    
\end{equation}
where all the components are $4 \times 4$ matrices in the direct product of spin and layer spaces. The diagonal components $\hat g_N=-\hat {\tilde g}_N=1$  are to be calculated in the normal state of the superconductor. $\hat f$ and $\hat {\tilde f}$ are the anomalous components  of the Green's function.  The resulting linearized equation for the anomalous Green's function $\hat f$ takes the form:
\begin{align}
\{i\omega_m -\bm {\hat h}\bm \sigma, \hat f\}+[\hat \mu+t_{z}^F\rho_x, \hat f]+i\bm v_F\bm \nabla \hat f=0 .
\label{f_linearized_eq}    
\end{align}
This equation describes the spatial behavior of the anomalous Green's function in the bilayer. The amplitude of $\hat f$ is determined by the boundary conditions with a superconducting lead and is not of interest here. Introducing a vector $\bm F = (\hat f_{11}, \hat f_{12}, \hat f_{21}, \hat f_{22})^T$, where $\hat f_{ij}$ are components of the anomalous Green's function $\hat f$ in the layer space (they are matrices $2 \times 2$ in the spin space), it is convenient to rewrite Eq.~(\ref{f_linearized_eq}) in the standard for a linear system form:
\begin{align}
    \hat A \bm F+i(\bm v_F \bm \nabla) \bm F = 0
\end{align}

\begin{widetext}
    \begin{align}
\hat A=\left(
\begin{array}{cccc}
-2\bm h_1 \bm \sigma+2i\omega_m & -t_{z}^F & t_{z}^F & 0 \\
-t_{z}^F & -(\bm h_1+\bm h_2)\bm \sigma+(\mu_1-\mu_2) +2i\omega_m & 0 & t_{z}^F \\
t_{z}^F & 0 & -(\bm h_1+\bm h_2)  \bm \sigma+(\mu_2-\mu_1)+2i\omega_m & -t_{z}^F \\
0 & t_{z}^F & -t_{z}^F & -2\bm h_2 \bm \sigma+2i\omega_m
\end{array}
\right)
\end{align}
In this work we only consider collinear magnetic orientations of $\bm h_1$ and $\bm h_2$. Then $\bm F$ is diagonal in spin space and takes the form $\bm F = \bm F_\uparrow (1+\sigma_z)/2 + \bm F_\downarrow (1-\sigma_z)/2$. The anomalous Green's functions $\bm F_{\uparrow}$ can be expressed as a sum of four solutions physically corresponding to different types of Cooper pairs, which can exist in the bilayer, with different character of the penetration into the nonsuperconducting region: 
\begin{align}
    \bm F=\sum_{k=1,2,3,4} C_k \hat w_{k} e^{(-\frac{2 |\omega_m|}{v_F} + i\lambda_k)l},
\end{align}
where $\hat w_k$ are eigenvectors containing the layer structure of the corresponding solution, $C_k$ is an amplitude of the solution, which depends on the details of the boundary conditions at the interface between the superconductor and the bilayer, $v_F = |\bm v_F|$ and $l>0$ is the coordinate along the trajectory determined by $\bm v_F$. The spatial behavior of the proximity-induced Cooper pairs consists of an overall decay at the scale $v_F/2 |\omega_m|$ superimposed by oscillations with the inverse period $\lambda_k/2 \pi$, where
\begin{align}
    \lambda_k=\frac{-(h_1+h_2)}{v_F}\pm \frac{1}{\sqrt{2}v_F}\sqrt{\Delta h^2+\Delta \mu^2+4t_{z}^{F2}\pm\sqrt{(\Delta h^2-\Delta \mu^2)^2+8(\Delta h^2+\Delta\mu^2)t_{z}^{F2}+16 t_{z}^{F4}}},
    \label{eq:periods}
\end{align}
\end{widetext}
where $\Delta h = h_1 - h_2$. $\bm F_\downarrow$ is obtained from $\bm F_\uparrow$ with the substitution $h_{1,2} \to -h_{1,2}$. Eigenvalues $\lambda_k$, which determine the oscillation periods of all 4 solutions, are presented in Fig.~\ref{fig:oscillations} as functions of $\Delta \mu = \mu_1 - \mu_2$ for different types of magnetic bilayers: ferromagnetic [Figs.~\ref{fig:oscillations}(a)-(c)], antiferromagnetic [(d)-(f)], bilayer consisting of a ferromagnetic and a paramagnetic layer [(g)-(i)] and a ferrimagnetic bilayer [(j)-(l)]. The difference of on-site energies of the monolayers $\Delta \mu$ can be varied by the gating potential applied to one of the monolayers. The different horizontal rows of Fig.~\ref{fig:oscillations} correspond to an increase of the interlayer coupling $t_z^F$ from top to bottom. 

\begin{figure}[tb]
	\begin{center}
		\includegraphics[width=85mm]{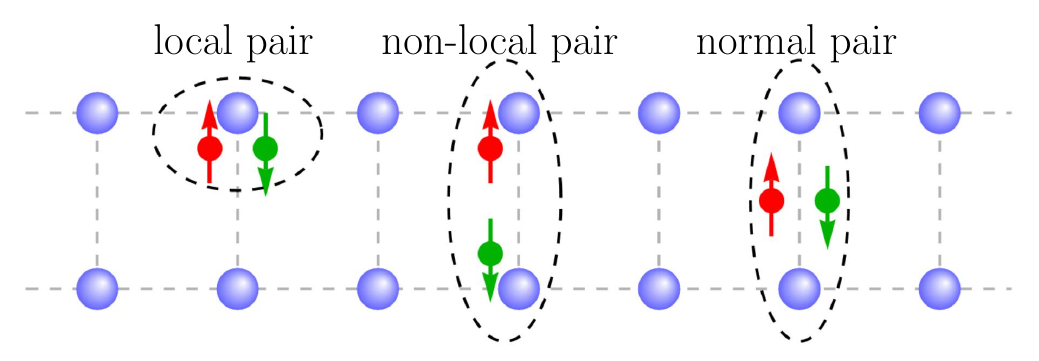}
\caption{Sketch of different types of proximity-induced Cooper pairs in the bilayer. The pairs can be classified as local and non-local at rather small values of the interlayer hopping $t_z^F$. At large values of the interlayer hopping the wave functions of paired electrons are homogeneously distributed over both layers and, thus, the pairs are called ``normal" because they are similar to the pairs in isotropic materials.}
 \label{fig:pairs}
	\end{center}
 \end{figure}

The upper row in Fig.~\ref{fig:oscillations}, representing the case of the lowest interlayer coupling, can be easily interpreted in terms of intralayer (local) and interlayer (non-local) Cooper pairs, where the paired electrons reside at the same monolayer or at different monolayers, respectively, see Fig.~\ref{fig:pairs} for sketch of such pairs. From Fig.~\ref{fig:oscillations}(a) it is seen that $\lambda_1 = \lambda_2 = -2h/v_F$ (where $h=h_1=h_2$). The corresponding eigenvectors are $\hat w_1 = (1,0,0,0)^T$ and $\hat w_2 = (0,0,0,1)^T$ up to the zero order with respect to small parameter $t_z^F$. That is, these correlations are local correlations residing at the first and second monolayers, respectively. Physically they represent the well-known mesoscopic Fulde-Ferrel-Larkin-Ovchinnikov (FFLO) state. Such oscillations of the Cooper pairs in the ferromagnet were predicted long ago \cite{Buzdin1982,Radovic2003} and lead to $0-\pi$ transitions in S/F/S JJs. However, two other proximity-induced pairing states corresponding to $\lambda_{3,4} = (-2h\pm \Delta \mu)/v_F$ and eigenvectors $\hat w_{3}=(0,1,0,0)^T$ and $\hat w_{4}=(0,0,1,0)^T$ represent non-local pairs consisting of electrons residing at different monolayers. Local pairs are not sensitive to $\Delta \mu$, which is natural because the paired electrons only feel the chemical potential of their own monolayer. Their oscillation period can depend on $\mu$ only via the dependence $v_F(\mu)$, but it requires much larger variations of the chemical potential to become pronounced here. At the same time the behavior of non-local pairs, which also represent finite-momentum FFLO-type pairing, is completely different. Their inverse oscillation period depends on $\Delta \mu$ linearly. The possibility to sustain such pairs, for which the oscillation period can be adjusted by gating, is a specific and very promising property of layered materials with a small number of atomic layers. Below we demonstrate how the gate-tunability of the non-local pairs opens perspectives of implementing the gate-controllable $0-\pi$ transitions in the appropriate JJs. 

Figs.~\ref{fig:oscillations}(d),(g) and (j) represent a generalization of the results shown in Fig.~\ref{fig:oscillations}(a) for the cases of unequal magnetizations of the monolayers composing the bilayer. At an arbitrary $h_1 \neq h_2$ there are two local oscillating proximity-induced pairing correlations with $\lambda_{1,2} = -2h_{1,2}/v_F$ corresponding to paired electrons residing at the first (second) monolayer, respectively. These pairing correlations are not adjustable by gating. The remaining pairing correlations correspond to non-local pairs having linear in $\Delta \mu$ inverse oscillation period $\lambda_{3,4} = [- (h_1+h_2) \pm \Delta \mu]/v_F$. In particular, in the antiferromagnetic case $h_1 = -h_2$ and at $\Delta \mu = 0$ the non-local pairs in the bilayer do not oscillate.

As the interlayer hopping increases, local and non-local pairs hybridize and the eigenstates of the proximity pair correlations become their mixture. This case is shown in the middle row of Fig.~\ref{fig:oscillations}. All four states are sensitive to gating because the eigenvectors of all the states have non-local components. With further increase of interlayer hopping, the spatial behavior of the proximity-induced Cooper pairs evolves towards the case of isotropic 3D metal, where it is impossible to introduce the concept of local pair, since the size of the Cooper pair in the direction of the $z$-axis significantly exceeds the interlayer distance. The case of the largest interlayer hopping considered is presented in the bottom row of Fig.~\ref{fig:oscillations}. Although the corresponding value of the interlayer hopping is still significantly less than the intralayer hopping, i.e. the system under consideration is far from the case of the isotropic material, the evolution of spatial behavior of Cooper pairs is already clearly traced. Indeed, we can see that the dependence of the oscillation period of all Cooper pairs on $\Delta \mu$ becomes weaker and gradually disappears. The reason is that in the limit of large $t_z^F \gg (h_{1,2},\Delta \mu)$ two eigenstates correspond to conventional Cooper pairs with $\lambda_{1,2} \approx -(h_1+h_2)/v_F$. The eigenvector of such pairs is $\hat w_{1,2} \approx (1,\pm 1,\pm 1,1)^T$, that is the wave functions of electrons making the pair are homogeneously distributed over both monolayers, forming a ``normal pair" [see Fig.~\ref{fig:pairs}]. Physically it is obvious that electrons of this pair feel the on-site energies of both monolayers equally and for this reason it is insensitive to $\Delta \mu$, as it is well-known for such pairs in isotropic systems. The other two pairs have a very rapidly oscillating wave function with $\lambda_{3,4} \approx \pm 2t_z^F/v_F$ and $\hat w_{3,4} = (-1,\pm 1,\mp 1,1)^T$.

\section{Josephson current through magnetic vdW bilayers}

\label{Josephson}

Here we calculate the Josephson current through the vdW magnetic bilayer in the framework of the model considered in Sec.~\ref{system}. Our main goal is to investigate the role of the non-local pair contribution: its relative magnitude with respect to the contribution of local pairs, specific manifestations in the Josephson current and significance for applications resulting from their sensitivity to gating. The calculations are performed exploiting the formalism of Bogolubov-de Gennes (BdG) equations.  

We diagonalize Hamiltonian (\ref{eq:hamiltonian}) by the Bogoliubov transformation:
\begin{align}
c_{\bm i\sigma}=\int \frac{d p_y}{2\pi}e^{i p_y i_y}\left(\sum\limits_n u_{n\sigma}^{\bm \eta}( p_y)\hat b_n+v^{\bm \eta *}_{n\sigma}( p_y)\hat b_n^\dagger \right) , 
\label{bogolubov}
\end{align}
where translation invariance in the $y$-direction is taken into account and $\bm \eta=(i_x, 0,i_z)^T$ is the projection of $\bm i$ onto the $xz$ plane.
Then the resulting Bogoliubov – de Gennes equations take the form:
\begin{align}
\xi_{\bm \eta}(p_y)u_{n,\sigma}^{\bm \eta} + \sigma \Delta_{\bm \eta} v^{ \bm \eta}_{n,-\sigma}- \\ \nonumber
- \sum\limits_{< \bm \eta \bm \eta'>} t_{\bm \eta \bm \eta'} u^{\bm \eta}_{ n, \sigma}+(\bm h_{\bm \eta} \bm{\sigma})_{\sigma\alpha}u_{n,\alpha}^{\bm \eta} & = \varepsilon_n u_{n,\sigma}^{\bm \eta} \nonumber \\  
\xi_{\bm \eta}^\sigma(p_y)v_{n,\sigma}^{\bm \eta} + \sigma \Delta_{ \bm \eta}^* u^{\bm \eta}_{n,-\sigma}- \\ \nonumber
-\sum\limits_{< \bm \eta \bm \eta'>} t_{\bm \eta \bm \eta'} v^{\bm \eta}_{ n, \sigma}+(\bm h_{\bm \eta}\bm{\sigma}^*)_{\sigma\alpha}v_{n,\alpha}^{\bm \eta} & = -\varepsilon_n v_{n,\sigma}^{\bm \eta}, 
\label{bdg}
\end{align}
where $\xi_{\bm \eta}(p_y)=-2t_{\bm \eta,\bm \eta+\bm a_y}\cos(p_ya)$.

The current between sites $\bm \eta$ and $\bm \eta'$ per unit length along the $y$-axis can be calculated via the solutions of the Bogoliubov-de Gennes equation as follows:
\begin{align}
    \bm j_{\bm \eta \to\bm\eta'}=e\sum_{n,\sigma}\int \frac{dp_y}{2\pi} it_{\bm \eta \bm \eta'}  \left((u_{n\sigma}^{ \bm \eta} u_{n\sigma}^{\bm \eta'*}-u_{n\sigma}^{ \bm \eta'} u_{n\sigma}^{\bm \eta*})f_n \right. \nonumber \\  
    \left.+(v_{n\sigma}^{ \bm\eta*} v_{n\sigma}^{\bm \eta'}-v_{n\sigma}^{ \bm\eta'*} v_{n\sigma}^{\bm \eta})(1-f_n)\right).
\end{align}

\subsection{Nonlocal contribution to Josephson current through AF bilayers}

\label{nonlocal_JJ}

\begin{figure}[tb]
	\begin{center}
		\includegraphics[width=87mm]{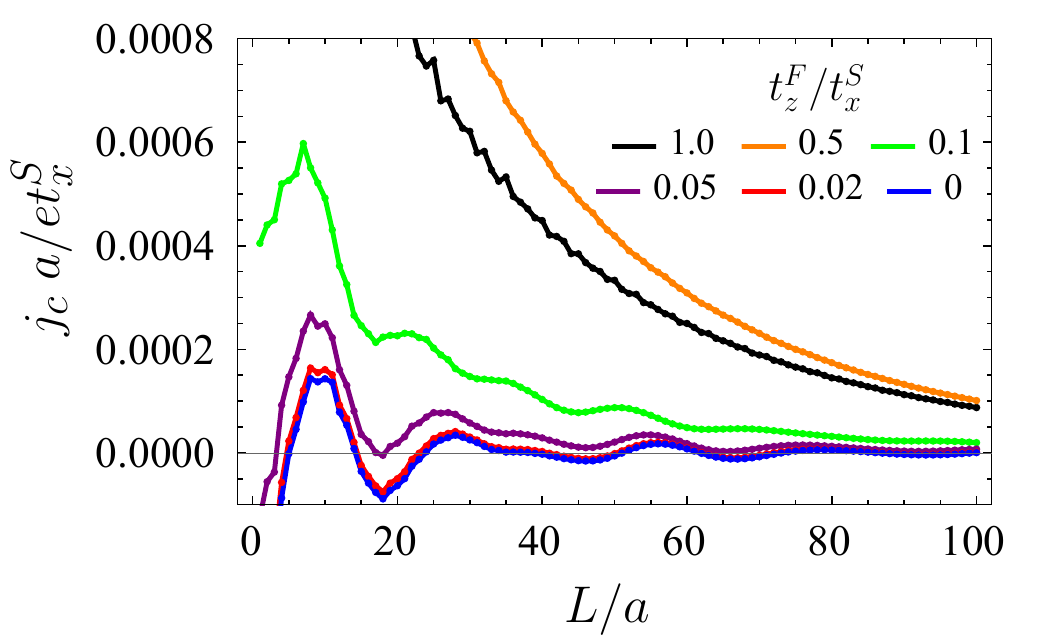}
\caption{Critical Josephson current as a function of the magnetic bilayer length $L$ for the case of the antiferromagnetic bilayer with $h_1 = -h_2 = 0.25$. No gating is assumed, that is $\Delta \mu = 0$. $L_h=16a, L_s=10a, t^F_x=t^S_z=t^S_x$, $t^{SF}=0.1t^S_x$, $T=0.07\Delta$, $\Delta=0.1t^S_x$. }
 \label{fig:current_AF_length}
	\end{center}
 \end{figure}

At first we present the results for the Josephson current via the AF bilayer with $h_1=-h_2=h$ assuming equal on-site energies of both magnetic layers (no gating). The critical Josephson current $j_c$ as a function of the length $L$ of the AF interlayer is demonstrated in Fig.~\ref{fig:current_AF_length}. $j_c$ is defined as the value of the Josephson current $j(\chi)$ at $\chi>0$ with maximal amplitude, which can be positive ($0$-JJ) or negative ($\pi$-JJ). Here $\chi$ is the superconducting phase difference between the leads. Different curves correspond to different values of the interlayer hopping $t_z^F$. It is seen that at low $t_z^F$ the critical current manifests oscillations. At such low value of the interlayer hopping the Josephson current is mainly transferred by local Cooper pairs. The oscillations of the critical current originate from the oscillations of the local Cooper wave function in the ferromagnetic layers. If $\Delta \mu = 0$ then the oscillation period of the local pairs is determined as $L_l = 2\pi \xi_{F,l}$ with $\xi_{F,l} = v_F /2\sqrt{(h^2 + t_z^{F2})}$.    The Josephson current is carried by the pairs traveling along different trajectories. It is known that for ferromagnetic interlayers containing many atomic layers and at $\xi_F \ll L \ll \xi_N$ the oscillations of the critical current are superimposed by the power law decay $j_c \sim (L/\xi_F)^{(1-n)/2}$, where $n$ is the dimensionality of the space \cite{Buzdin1982,Konschelle2008}. The decay is due to   averaging of the contributions corresponding to different trajectories.  The local pair contribution in our case behaves in the same way with the substitution $\xi_F \to \xi_{F,l}$. However, in the numerical calculations we do not restrict ourselves by the limit $L \ll \xi_N$ and for this reason this power law decay is further superimposed by the exponential decay $\sim e^{-L/\xi_N}$ \cite{Golubov2004}. 

As the interlayer hopping increases, a non-oscillating contribution is added to the oscillating  Josephson current, which decays at the normal coherence length and becomes dominant at large values of $t_z^F$. This contribution is due to the nonlocal pairs, which do not oscillate in the AF bilayer at $\Delta \mu = 0$ according to the consideration presented in the previous section. The amplitude of such pairs contains an additional factor $t_z^F$ with respect to the amplitude of the local pairs and, therefore, their relative contribution to the Josephson current is $ \sim t_z^{F2}$. Due to the growing contribution of the non-local pairs the $0-\pi$ transitions in the JJ disappear with increase of $t_z^F$. Although at first glance the considered system physically look similar (except for the thickness of the layers) to S/F/S JJs with a magnetic interlayer composed of two parallel thick ferromagnetic layers with opposite magnetizations \cite{Buzdin2011}, the predicted supercurrent behavior is fundamentally different. It was demonstrated that if the AF layer is composed of two parallel oppositely magnetized thick F layers, the interface between the layers (domain wall) provides an additional channel for the Josephson current with effectively reduced dimension \cite{Buzdin2011}. In the ballistic case the additional contribution to the Josephson current was found to decay as $\delta j_c \sim (L/\xi_F)^{(2-n)/2}$ for $n=2,3$. This reduced decay was superimposed by standard oscillation with the period $2\pi \xi_F$. In our case, the contribution of non-local pairs does not oscillate at all, since they represent a completely different physical phenomenon. Namely, they should be interpreted as pair correlations of electrons with opposite spins, belonging to different layers, and not as an additional channel for the Josephson current of reduced dimensionality, where, nevertheless, both paired electrons feel the same exchange field.

At the largest values of $t_z^F$ when the anisotropy between the intralayer and interlayer hopping already disappears (this case is represented by the curve corresponding to $t_z^F = 1$ in Fig.~\ref{fig:current_AF_length}), the difference between local and non-local pairs is erased, as discussed in the previous section. In this case, from the point of view of the proximity effect our layered antiferromagnetic metal is practically no different from a normal metal.

\subsection{Gate-tunable $0-\pi$ transitions in JJs with AF interlayers}

\label{zero_pi}

As it was demonstrated in Sec.~\ref{nonlocal}, the oscillation period of the non-local Cooper pairs is sensitive to the difference $\Delta \mu$ between the on-site energies of both layers, in which the pair resides at. It is a specific property of the non-local pairs, which is not inherent to the local pairs and also disappears in the limit of an isotropic metal, when wave functions of the electrons are distributed over several layers. The pair oscillations manifest themselves as oscillations of the critical Josephson current. For this reason the sensitivity of the proximity pairs to $\Delta \mu$ results in the corresponding sensitivity of the critical current. Experimentally in vdW bilayers $\Delta \mu$ can be tuned by gating one of the layers. The possibility to use gating to tune on-site energy is an exciting property of vdW materials, which allows for a high degree of controllability of superconductivity \cite{Ueno2008,Ye2010,Zeng2018,Saito2016_SUST,Lu2018,Zhang2013},  ferromagnetism \cite{Deng2018,Matsuoka2023,Tan2021} and also magnetic proximity effect in nonsuperconducting \cite{Cardoso2023} and superconducting \cite{Bobkov2024_vdW,Bobkov2024_spin,Ianovskaia2024} vdW heterostructures. The dependence of the critical current on the gating potential can not only experimentally prove the presence of the non-local proximity-induced Cooper pairs, but is also of great interest for applications, as it results in the possibility of gate-tunable $0-\pi$ transitions, as demonstrated below.

\begin{figure}[tb]
	\begin{center}
		\includegraphics[width=85mm]{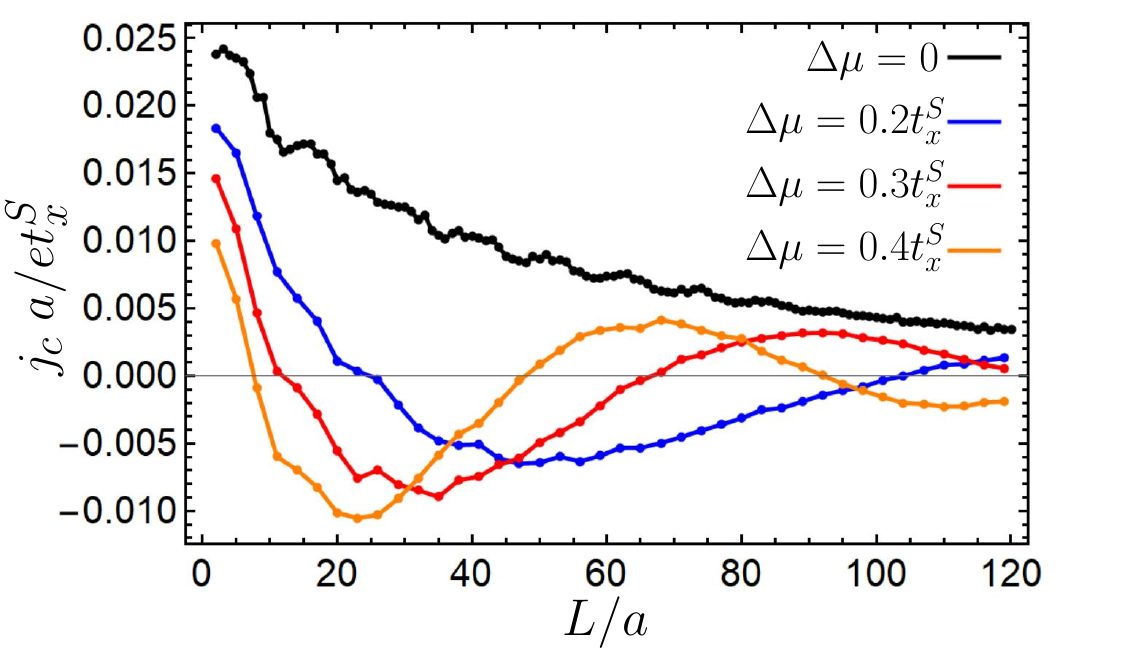}
\caption{Critical Josephson current as a function of the magnetic bilayer length $L$ for the case of the antiferromagnetic bilayer with $h_1 = -h_2$. Different curves correspond to different values of $\Delta \mu$. $t_z^F = t_z^{SF}=t_z^S=0.5t_x^F=0.5t_x^S$, $h_1=0.2t_x^S,\Delta=0.04t_x^S,T=0.075\Delta,\mu_S=\mu_{1,2}(\Delta\mu=0)=0.5t_x^S,L_h=6a,L_S=10a$.}
 \label{fig:AF_L_gating}
	\end{center}
 \end{figure}

The dependence of the critical current on $L$ for the case of the AF interlayer and different values of $\Delta \mu$ is presented in Fig.~\ref{fig:AF_L_gating} for $t_z^F=0.5 t_x^F$. At $\Delta \mu = 0$ for this value of the interlayer hopping  the Josephson current does not manifest $0-\pi$ transitions. However, from Fig.~\ref{fig:AF_L_gating} one can observe that with increasing $\Delta \mu$, oscillations of the critical current on the length of the AF weak link and $0-\pi$ transitions appear. The period of the oscillations decreases with increasing $\Delta \mu$. The observed behavior of the critical current can be explained by the fact that at $t_z^F = 0.5 t_x^F$ the local and nonlocal pairs are already strongly hybridized and for this reason the oscillation periods essentially depend on $\Delta \mu$.

From Eq.~(\ref{eq:periods}) at $t_z^F \gg (\Delta \mu, h)$ and $h_1 = -h_2 = h$ one can obtain $\lambda_{1,2} \approx \mp h \Delta \mu/v_F t_z^F$ with $\hat w_{1,2} \approx (1, \pm 1, \pm 1, 1)^T$  and $\lambda_{3,4} \approx \mp [2t_z^F + (h^2 + \Delta \mu^2/4)/v_F t_z^F]$ with $\hat w_{3,4} \approx (-1, \mp 1, \pm 1, 1)^T$.  At $\Delta \mu = 0$ the main contribution to the Josephson current is provided by the pairs corresponding to $\lambda_{1,2}$ with no spatial oscillations. The contribution of the pairs corresponding to $\lambda_{3,4}$ is also seen as small oscillations, accompanying the overall decay. The corresponding oscillation period, obtained from the results of the BdG calculations, is equal to $L_{BdG}=11a$, which is in a good agreement with the analytical formula $\frac{2\pi}{|\lambda_{3,4}|}=13a$. Yet the amplitudes of $\lambda_{3,4}$ pairs are small as compared to $\lambda_{1,2}$ pairs. The reason is that the eigenvectors $\hat w_{1,2}$ are even with respect to the interchange of both magnetic layers $1 \leftrightarrow 2$ (even layer parity) and the eigenvectors $\hat w_{3,4}$ are odd with respect to such interchange (odd layer parity). The last type of pairs should not be generated at an interface with a fully symmetric two-layer superconductor. We believe that in our case a small admixture of such pairs is generated at the interface with the superconductor due to the breaking of the layer symmetry because the superconducting leads are in direct contact only with the upper magnetic layer. At $\Delta \mu \neq 0$ the $\lambda_{1,2}$ pair become oscillating and for this reason the corresponding current contribution manifests a $0-\pi$ transition as a function of the interlayer length. The behavior of the current in Fig.~\ref{fig:AF_L_gating} is in good agreement with the above approximate formulas. Namely, for $\Delta\mu=0.2,0.3,0.4t_x^S$ the periods $L_{BdG}\approx172a,112a,85a$ are obtained from the numerical results of the BdG calculations, while the above formula gives $\frac{2\pi}{|\lambda_{1,2}|}=169a,112a,84a$. The small admixture of the $\lambda_{3,4}$ pairs is still seen in the curves corresponding to $\Delta \mu \neq 0$.

\begin{figure}[tb]
	\begin{center}
		\includegraphics[width=85mm]{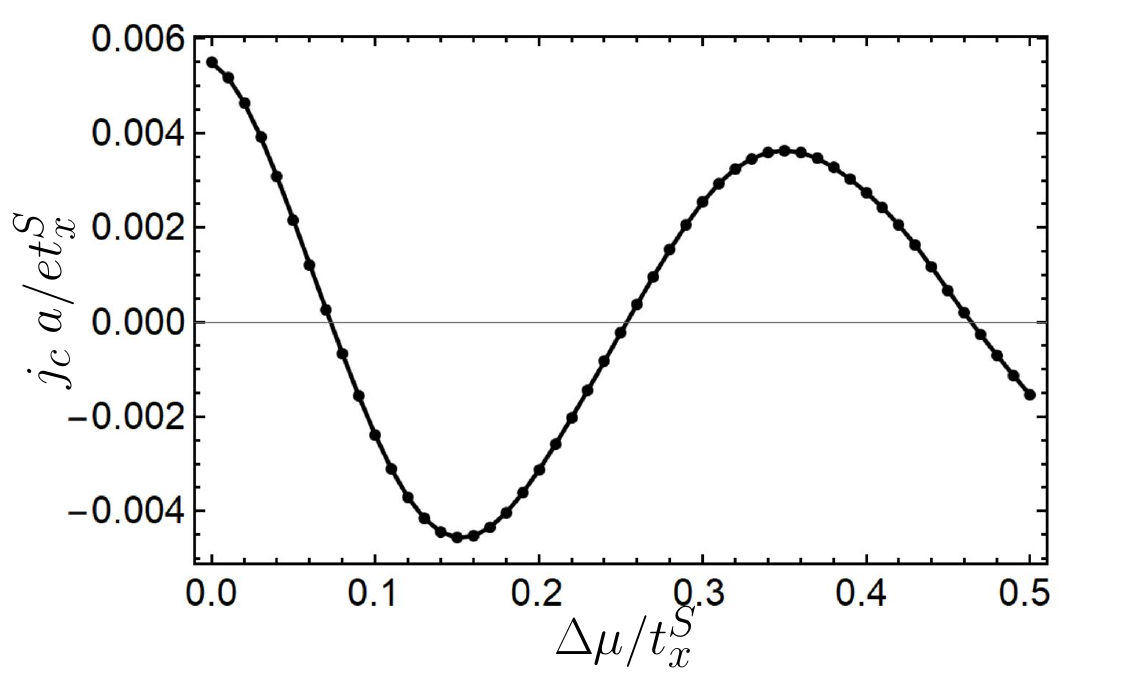}
\caption{AF bilayer. Critical Josephson current as a function of the difference between on-site energies of magnetic layers $\Delta \mu$. $L=80a$, other parameters are the same as in Fig.\ref{fig:AF_L_gating}.}
 \label{fig:AF_gating}
	\end{center}
 \end{figure}

The strong sensitivity of the oscillation period of the pairs to $\Delta \mu$ provides a possibility to implement $0-\pi$ transitions upon varying the gating potential. It is demonstrated in Fig.~\ref{fig:AF_gating}, where the critical current is plotted as a function of $\Delta \mu$ for a given length of the AF interlayer.

\subsection{Josephson current through F interlayers}

\label{ferromagnet_JJ}

\begin{figure}[tb]
	\begin{center}
		\includegraphics[width=85mm]{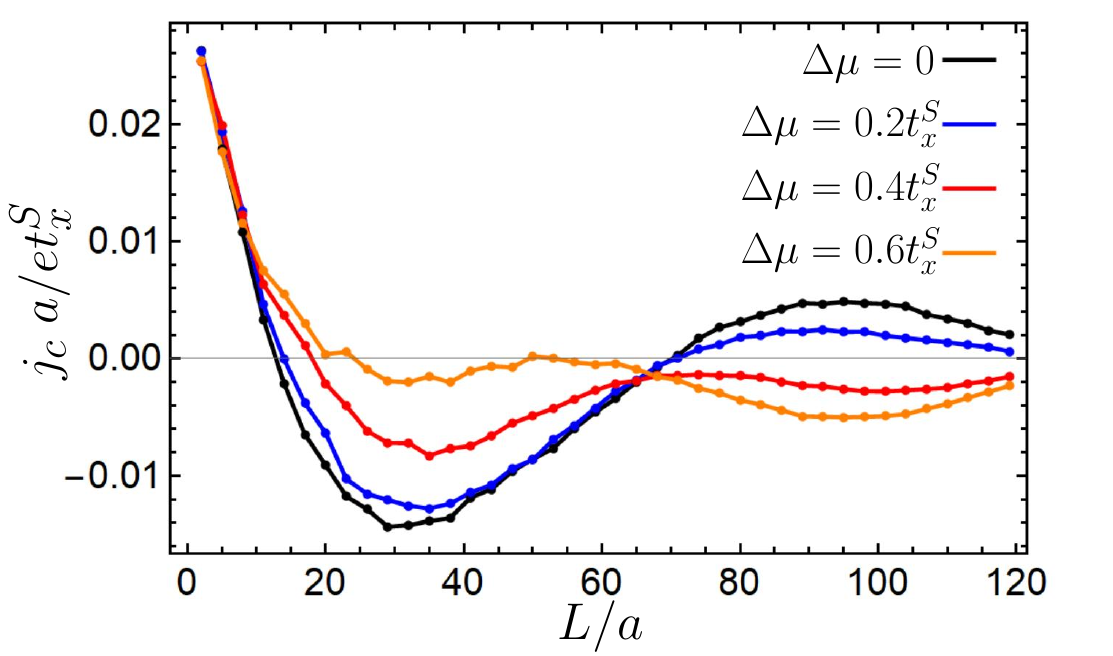}
\caption{Critical Josephson current as a function of the magnetic bilayer length $L$ for the case of the  bilayer composed of a ferromagnetic and a normal layers with $h_1 = h$ and $h_2=0$. Different curves correspond to different values of $\Delta \mu$. $h_1=0.1t_x^S$, other parameters are the same as in Fig.\ref{fig:AF_L_gating}.}
 \label{fig:NF_L_gating}
	\end{center}
 \end{figure}

The case of the AF bilayer considered in the previous section is special in the sense that at $\Delta \mu = 0$ there are no oscillations of the critical current due to $h_1 = -h_2$. Then the sensitivity of the system to gating is the strongest. If the condition $h_1 = -h_2$ is violated the critical current begins to oscillate as a function of the interlayer length. As an example we consider an interlayer composed of a ferromagnetic and normal layers with $h_2=0$. The critical current as a function of the interlayer length is plotted in Fig.~\ref{fig:NF_L_gating} for different values of $\Delta \mu$. It is seen that the critical current oscillates even at $\Delta \mu = 0$. The oscillation period is in good agreement with analytical results, which  at $t_z^F \ll (\Delta \mu, h)$ can be expanded as $\lambda_{1,2} \approx -h/v_F\mp h\Delta \mu /2v_F t_z^F$ with $\hat w_{1,2} \approx (1, \pm 1, \pm 1, 1)^T$, which provide the leading contribution to the Josephson current, and $\lambda_{3,4} \approx (\mp 2t_z^F-h)/v_F \mp (h^2 + \Delta \mu^2)/4 v_F t_z^F$ with $\hat w_{3,4} \approx (-1, \mp 1, \pm 1, 1)^T$, which only make a minor contribution to the critical current. 

\begin{figure}[tb]
	\begin{center}
		\includegraphics[width=85mm]{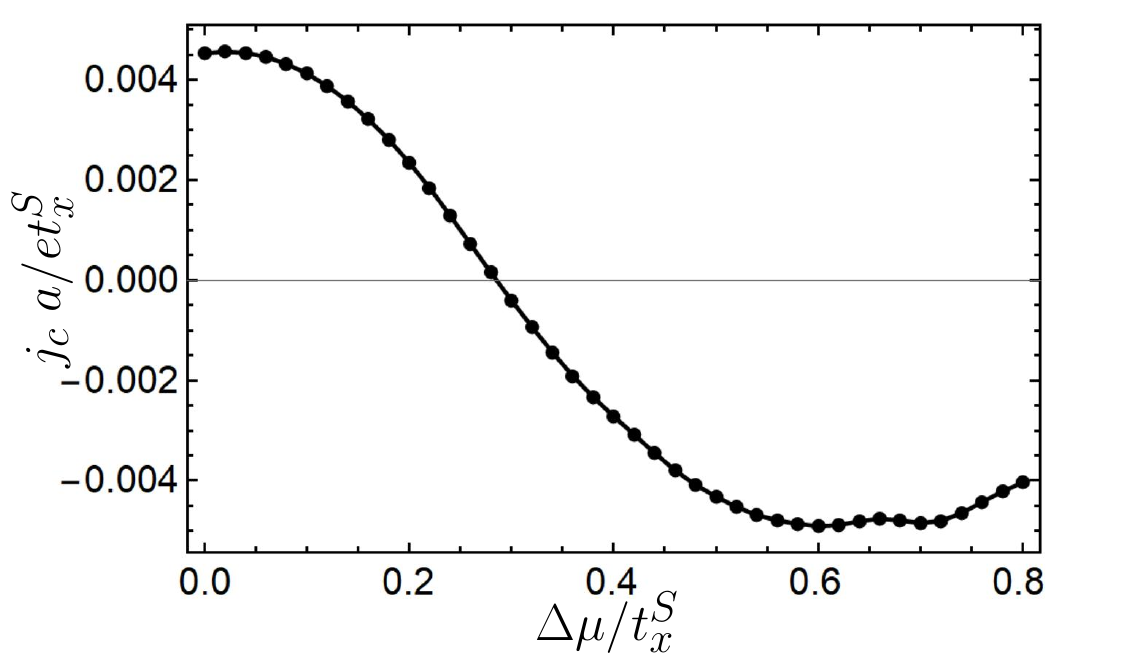}
\caption{Bilayer composed of a ferromagnetic and a normal layers. Critical Josephson current as a function of the difference between on-site energies of magnetic layers $\Delta \mu$. $L=100a$, other parameters are the same as in Fig.\ref{fig:NF_L_gating}.}
 \label{fig:NF_gating}
	\end{center}
 \end{figure}

In the considered case for small $\Delta \mu$ the dependence of the oscillation period on $\Delta \mu$ is only a correction to the main term, which is associated to $\lambda_{1,2} \approx -h/v_F$. At large $\Delta \mu \gg t_z^F$ the difference between on-site energies effectively works as an additional  barrier between the layers (mismatch effect). Then it is obvious that in this case the Josephson current is just a sum of independent contributions of two individual layers, one of which is oscillating due to the presence of $h_1=h$ and the other one is non-oscillating due to $h_2=0$. Thus, even in this case one can control $0-\pi$ transitions by applying the gating voltage, see Fig.~\ref{fig:NF_gating}. In general, the Josephson current is sensitive to gating for a wide range of magnetic bilayers except for the case of a ferromagnetic bilayer with $h_1 = h_2$ because in this case the oscillation periods of the pairs, which provide the leading contribution to the Josephson current, do not depend on $\Delta \mu$, as  seen from the left column of Fig.~\ref{fig:pairs}. 

\section{Conclusions}

\label{conclusions}

In this paper a theory of proximity effect in ballistic bilayer vdW magnetic systems is developed. Possible types of proximity-induced Cooper pairs are studies and their spatial behavior is described analytically. The considered proximity effect in the vdW bilayer systems is shown to be uniquely characterized by the presence of non-local Cooper pairs, whose spatial behavior in the non-superconducting regions is qualitatively different from the well-known proximity effect in conventional non-magnetic and magnetic non-superconducting materials. 
In particular, the oscillation period of spatial oscillations of such pairs in non-superconducting regions is shown to be  sensitive to the difference between on-site energies of the monolayer composing the bilayer and, thus, can be controlled by applying a gating potential to one of the monolayers. This opens the perspective of implementing the gate-controlled $0-\pi$ transitions in Josephson junctions through weak links made of such bilayers.

\begin{acknowledgments}
G.A.B., D.S.R. and I.V.B. acknowledge the support from Theoretical Physics and Mathematics Advancement Foundation “BASIS” via the project No. 23-1-1-51-1. The analytical calculations in the framework of the quasiclassical approach were supported by the Russian Science Foundation via the RSF project No. 24-12-00152. The numerical calculations of the Josephson current were supported by Grant from the ministry of science and higher education of the Russian Federation No. 075-15-2024-632.
\end{acknowledgments}

\bibliography{JJ_layeredAF}

\end{document}